\documentclass[twocolumn,showpacs,preprintnumbers,amsmath,amssymb]{revtex4-1}
\usepackage{graphicx}
\usepackage{textcomp}

\begin{document}
\title{Quantum decay and amplification in a non-Hermitian unstable continuum}
  \normalsize
\author{Stefano Longhi}
\address{Dipartimento di Fisica, Politecnico di Milano and Istituto di Fotonica e Nanotecnologie del Consiglio Nazionale delle Ricerche, Piazza L. da Vinci
32, I-20133 Milano, Italy}

%
\bigskip
\begin{abstract}
The decay of a bound state weakly-coupled  to a non-Hermitian tight-binding unstable continuum, i.e. a continuum of states comprising energies with positive imaginary part, is theoretically investigated.  As compared to quantum decay in an Hermitian continuum, in the non-Hermitian case a richer scenario can be found as a result of non-unitary dynamics.
Different behaviors are observed depending on the kind of instability of the continuum. These include complete or fractional decay in convectively-unstable continua, the absence of quantum decay for a bound state with energy embedded in the continuum loop, and unstable (secular) growth with pseudo exponential amplification in the absolutely-unstable regime. Analytical results are presented for a nearest-neighboring tight-binding continuum with asymmetric hopping rates $\kappa_1$ and $\kappa_2$, which shows a transition from convective to absolute instability when the sign $\kappa_1 \kappa_2$ changes from positive to negative. In the convectively unstable regime the model describes the decay of a bound state coupled to a tight-binding lattice with an imaginary gauge field, which shows a pseudo-Hermitian dynamics.  In the absolutely-unstable regime, pseudo-Hermitian dynamics is broken and a pseudo exponential secular growth is observed. 

\noindent

\end{abstract}




\maketitle

\section{Introduction}

The decay of unstable states into a continuum is commonplace in many areas of physics, ranging from quantum physics \cite{r1,r2,r3} to statistical mechanics \cite{r4,r5}, atomic and molecular physics \cite{r6,r6bis,r6tris}, optics \cite{r7,r8} and cosmology \cite{r9}.  While an exponential decay is ubiquitous in nature, quantum mechanics dictates that the decay law should deviate from an exponential one at short and long time scales. Such deviations may lead to deep physical implications. Short-time deviations of the decay process have been observed in experiments on macroscopic quantum tunneling of cold atoms \cite{r10} and have attracted a considerable interest because of the possibility to either decelerate
(Zeno effect) or accelerate (anti-Zeno effect) the decay by frequent observations of the system \cite{r11,r12,r13,r14,r15,r16,r17}. At long time scales the quantum mechanical decay slows down and shows a power-law decay \cite{r17bis}. Slowing down the decay process may have major implications in cosmological models \cite{r9}, for example it may increase the likelihood of eternal inflation. In many particle systems, the decay dynamics can be modified by particle statistics and contact interactions \cite{r18}. Recent experiments demonstrated the role of Pauli exclusion principle \cite{r19} and interaction-induced particle correlations \cite{r20}. \par In such previous studies, quantum mechanical decay has been mostly considered in the framework of standard quantum mechanics, which sets the Hamiltonian $\hat{H}$ of the full system to be an Hermitian operator. 
Recently, an increasing interest has been devoted to investigate the dynamics of non-Hermitian quantum and classical systems \cite{r21}, especially those possessing parity-time ($\mathcal{PT})$ symmetry \cite{r22}. Such systems find important applications in several areas of physics. For example, effective non-Hermitian Hamiltonians are often introduced in the description of open quantum systems. In optics, non-Hermitian and $\mathcal{PT}$-symmetric dynamics can be exploited to the design of integrated photonic devices with novel functionalities \cite{r23}. Such studies motivate to extend the standard quantum mechanical description of the decay process to the non-Hermitian case, where the non-unitary dynamics is expected to modify the decay process or to induce an unstable growth (rather than decay) when the energy spectrum shows energies with positive imaginary part. Recent studies have considered the decay problem of discrete states with complex energies coupled to an Hermitian continuum, predicting the existence of bound states either outside or embedded into the continuum \cite{r24}. However, the more general case of the decay into a {\it non-Hermitian continuum} has been so far overlooked. Noteworthy examples of non-Hermitian continua are provided by so-called complex crystals, i.e. periodic complex potentials, which can be experimentally implemented in optics and in cold atom systems \cite{r25,r26}. Complex crystals have recently attracted a great attention because of their rather unique scattering properties, such as the ability to appear invisible when probed on one side \cite{r27}, to realize Talbot self-imaging \cite{r28} and to show a giant Goos-H\"{a}nchen shift \cite{r29}.\par
In this work we theoretically investigate the decay dynamics of a Hermitian bound state of (real) energy $\omega_a$ weakly coupled to a tight-binding {\it non-Hermitian} continuum [Fig.1(a)]. The non-Hermitian continuum is realized by a tight-binding (single-band) crystal with a complex energy dispersion curve $\omega=\omega(k)$, that describes a closed loop $\mathcal{L}$ in the complex energy plane when the Bloch wave number $k$ spans the Brillouin zone. The continuum is assumed to be {\it unstable}, i.e. the loop $\mathcal{L}$ invades the $\rm{Im}(\omega)>0$ half complex energy plane [Fig.1(b,c)]. In the limiting case of a Hermitian continuum and in the weak coupling limit, i.e. for $\omega(k)$ real and the loop $\mathcal{L}$ shrinking to a line on the real axis ${\rm Im} (\omega)=0$, as a general rule it is well-known that the decay into the continuum is complete whenever the energy $\omega_a$ of the discrete state is embedded into the tight-binding energy band, whereas the decay is limited (fractional) when $\omega_a$ falls outside the tight-binding energy band. Such a rule can fail in very special cases, corresponding to the existence of bound states embedded in the continuum. What happens when the discrete state of energy $\omega_a$ is weakly coupled to an unstable non-Hermitian tight-binding continuum?  As we will show in the present work, a richer scenario can be found, mainly depending on the instability properties of the continuum. For a so-called {\it convectively-unstable} continuum \cite{r30}, complete quantum decay can persist in spite of energy states with positive imaginary part in the continuum, however contrary to the Hermitian case incomplete (fractional) decay can be found  even when the energy $\omega_a$ is embedded into the continuum, i.e. $\omega_a$ falls inside the loop $\mathcal{L}$. For an {\it absolutely-unstable} continuum \cite{r30}, a secular growth (rather than decay) is observed, however contrary to a naive prediction the amplification is not purely exponential, rather it is pseudo exponential. The general analysis is exemplified by considering a nearest-neighboring tight-binding continuum with asymmetric hopping rates, which is described by the dispersion relation $\omega(k)=\Delta_1 \cos k+i \Delta_2 \sin k$. The continuum is convectively unstable for $\Delta_2<\Delta_1$, whereas it is absolutely unstable for $\Delta_2>\Delta_1$. In the former case it is shown that the quantum decay dynamics is pseudo-Hermitian, i.e. it can be reduced to the one of an equivalent Hermitian Hamiltonian via an 'imaginary' gauge transformation, whereas in the latter case the dynamics is non-Hermitian in its essence and a pseudo exponential secular growth is observed.

 \begin{figure}
\includegraphics[scale=0.3]{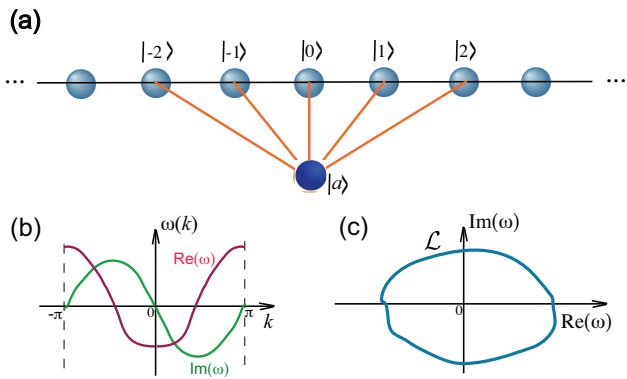}
\caption{(Color online) (a) Schematic of a discrete state $|a \rangle$ side coupled to a non-Hermitian tight-binding continuum (linear chain with Wannier states $|n \rangle$, $n=0, \pm 1, \pm 2, ...$). (b) Dispersione relation $\omega=\omega(k)$ of the unstable tight-binding continuum. Bloch modes at wave number $k$  with positive imaginary part of $\omega(k)$ are unstable modes. (c) Closed path $\mathcal{L}$ followed by the complex energy $\omega(k)$ when the Bloch wave number $k$ spans the Brillouin zone.}
\end{figure}

\section{Quantum decay/amplification of a discrete state coupled to an unstable tight-binding continuum: general analysis}
\subsection{The model}
Let us consider a discrete state $|a \rangle$ with real energy $\omega_a$, which is side-coupled to a one-dimensional non-Hermitian tight-binding lattice (a quantum wire) with a {\it complex} energy dispersion curve $\omega=\omega(k)$ [Figs.1(a) and (b)]. In the Wannier basis representation $|n \rangle$ of the tight-binding lattice, the state vector  $|\psi(t) \rangle$ of the system can be expanded as
\begin{equation}
 |\psi(t) \rangle= c_a(t) |a \rangle +\sum_{n=-\infty}^{\infty} c_n(t) |n \rangle,
\end{equation}
  where the amplitude probabilities $c_a(t)$ and $c_n(t)$ satisfy the coupled equations
\begin{eqnarray}
i \frac{dc_a}{dt} & = & \omega_a c_a+ \sum_{-\infty}^{\infty} \sigma_n c_n \\
i \frac{dc_n}{dt} & = & \sum_{m} \omega_{n-m}c_m+\rho_n c_a.
\end{eqnarray}
In Eqs.(2) and (3), $\omega_l$ are the Fourier coefficients of the band dispersion curve, i.e.
\begin{equation}
\omega(k)=\sum_{l=-\infty}^{\infty} \omega_l \exp(ikl),
\end{equation}
$k$ is the Bloch wave number that varies in the first Brillouin zone $ -\pi \leq k < \pi$, and $\rho_n$, $\sigma_n$ describe the couplings between the discrete state $| a \rangle$ and the Wannier state $|n \rangle$ of the lattice. In the following, we will assume  Hermitian coupling, so that 
\begin{equation}
\rho_n= \sigma_n^*.
\end{equation}
  Since the dispersion relation $\omega(k)$ is a periodic function of the Bloch wave number $k$, as $k$ spans the Brilluoin zone, from $k=-\pi$ to $k= \pi$, $\omega=\omega(k)$ spans a closed loop $\mathcal{L}$ in the complex plane. For the sake of simplicity, we will assume that the closed path $\mathcal{L}$ is a single loop [Fig.1(c)], i.e. that $\omega(k_2) \neq \omega(k_1)$ for $k_2 \neq k_1$ and $(d \omega / dk)$ non singular. We also assume that the tight-binding continuum is {\it unstable}, i.e. there are energies $\omega(k)$ in the $\rm{Im}( \omega)>0$ complex plane. Typically, we assume energies in the $\rm{Im}( \omega)<0$ complex plane as well with balanced dissipation and amplification, i.e. 
  \begin{equation}
  \int_{-\pi}^{\pi} dk {\rm Im}[\omega(k)]=0
  \end{equation}
  which implies ${\rm Im}(\omega_0)=0$.  
  Following the definitions of unstable flows in hydrodynamic
systems \cite{r31} (see also Ref.\cite{r30}), the continuum is said to be {\it convectively unstable} if, for any initial localized excitation of the continuum $c_n(0)$, one can find a drift velocity $V \neq 0$ such that $|c_{n-Vt}(t)|$ is unbounded as $ t \rightarrow \infty$, but $c_{n}(t) \rightarrow 0$ as $ t \rightarrow \infty$ at any fixed lattice site $n$. The continuum is said to be {\it absolutely unstable} if $|c_{n}(t)|$ is unbounded as $ t \rightarrow \infty$ at any fixed position $n$. As shown in the Appendix A, a continuum with ${\rm Im}[\omega(k)]>0$ for some $k$ is always convectively unstable, whereas it is also absolutely unstable whenever ${ \rm Im} [\omega(k_s)]>0$, where $k_s$ is the most critical saddle point in the complex $k$ plane of the dispersion relation $\omega(k)$, i.e.
  \begin{equation}
  \left( \frac{d \omega}{dk} \right)_{k_s}=0.
  \end{equation}
  The above definition of convectively/absolutely unstable continuum refers to the case where the discrete state $|a \rangle$ is not coupled to the continuum. Let us now consider a non-vanishing discrete-continuum coupling and let us assume that, at initial time $t=0$, the system is prepared in the discrete state, so that Eqs.(2) and (3) should be integrated with the initial conditions $c_a(0)=1$ and $c_n(0)=0$. Our aim is to provide some general results on the time evolution of the occupation probability of the discrete state, i.e. $P_a(t)=|c_a(t)|^2$. Since the continuum is unstable, the dynamics is not unitary and the "survival" probability $P_a(t)$ can be either bounded or unbounded as $ t \rightarrow \infty$. To determine the general behavior of the survival probability, it is worth introducing the Bloch basis $| k \rangle= (1/ \sqrt{2 \pi}) \sum_{n=-\infty}^{\infty} \exp(ikn) |n \rangle$ and to expand the state vector of the system as 
\begin{equation}
|\psi(t) \rangle=c_a(t) | a \rangle + \int_{-\pi}^{\pi} c(k,t) |k \rangle.
\end{equation}
 Taking into account that 
\begin{equation}
c(k,t)=\frac{1}{\sqrt{2 \pi}}  \sum_{n=-\infty}^{\infty} c_n(t) \exp(ikn), 
\end{equation}
and using Eqs.(2,3), one readily obtains the following coupled equations for the amplitude probabilities $c_a(t)$ and $c(k,t)$ in the Bloch basis 
\begin{eqnarray}
i \frac{dc_a(t)}{dt} & = & \omega_a c_a(t)+ \int_{-\pi}^{\pi} dk g_1(k) c(k,t) \\
i \frac{dc (k,t)}{dt} & = & \omega(k) c(k,t)+ g_2(k)  c_a(t)
\end{eqnarray}
where we have set
\begin{eqnarray}
g_1(k) & = & \frac{1}{\sqrt{2 \pi}} \sum_{n=-\infty}^{\infty} \sigma_n \exp(-ikn) \\
g_2(k) & = & \frac{1}{\sqrt{2 \pi}}  \sum_{n=-\infty}^{\infty} \rho_n \exp(ikn).
\end{eqnarray}
Note that, for Hermitian coupling one has $g_2(k)=g_1^*(k)$.  

 \begin{figure}
\includegraphics[scale=0.3]{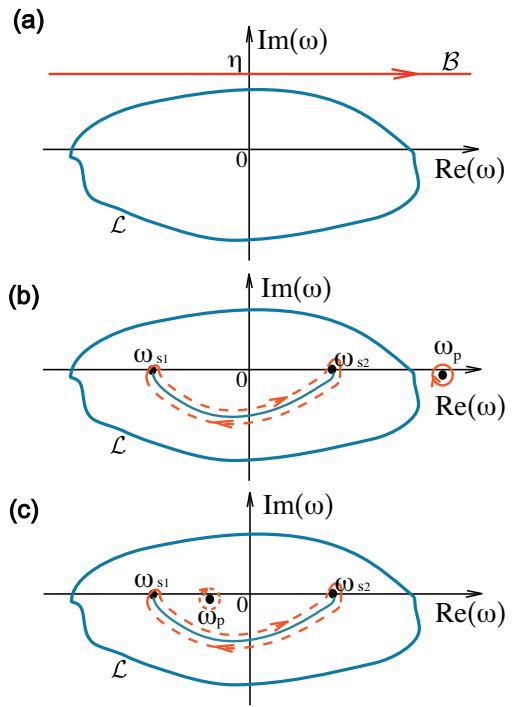}
\caption{(Color online) (a) Bromwich path $\mathcal{B}$ entering in the inverse Fourier-Laplace transform [Eq.(14)]. The closed loop $\mathcal{L}$ is a branch cut of the self-energy $\Sigma( \omega)$. (b) Deformation of the Bromwich path when $\omega_a$ falls outside the loop $\mathcal{L}$. $\omega_p$ is the pole of $\hat{c}_a(\omega)$ near $\omega_a$ outside the loop $\mathcal{L}$, whereas $\omega_s$ are the branch points of the density of states $\rho(\omega)$ internal to the loop $\mathcal{L}$. In the figure two branch points $\omega_{s1}$ and $\omega_{s2}$ are shown for the sake of definiteness. The solid line connecting the two branch points is the branch cut of $\hat{c}^{II}_a(\omega)$. The integral of $\hat{c}_{a}^{II}(\omega)$ along the dashed closed curve  that encircles the branch cut gives the branch cut contribution $c_a^{(cut)}(t)$ to $c_a(t)$, whereas the integral of $\hat{c}_{a}(\omega)$ over the solid circle around the pole $\omega=\omega_p$ gives the exponential term to $c_a(t)$ [see Eq.(23)]. (c) Same as (b), but when $\omega_a$ falls inside the loop $\mathcal{L}$. In this case the pole contribution arises from the integral of $\hat{c}_{a}^{II}(\omega)$ over the dashed circle around the pole $\omega=\omega_p$ near $\omega_a$.}
\end{figure}

\subsection{Quantum decay/amplification laws}
Likewise the quantum mechanical decay in the Hermitian case, the solution $c_a(t)$ to Eqs.(10) and (11) with the initial condition $c_a(0)=1$ and $c(k,0)=0$ can be conveniently
written as the inverse Fourier-Laplace transform of
the propagator \cite{r2,r14},  namely one has
\begin{equation}
c_a(t)=\frac{1}{2 \pi } \int_{\mathcal{B}} d \omega \; \hat{c}_a(\omega) \exp(-i \omega t)
\end{equation}
where 
\begin{equation}
\hat{c}_a(\omega)=\frac{1}{\omega-\omega_a-\Sigma(\omega)}
\end{equation}
is the Fourier-Laplace transform of $c_a(t)$ and 
\begin{equation}
\Sigma(\omega)=\int_{-\pi}^{\pi} dk \frac{g_1(k)g_2(k)}{\omega-\omega(k)}
\end{equation}
is the self-energy. In Eq.(14) the Bromwich path $\mathcal{B}$ is any horizontal line $\rm{Im}(\omega)=\eta$ in the complex $\omega$ plane which lies above the loop $\mathcal{L}$; see Fig.2(a). The self-energy $\Sigma(\omega)$ is not defined on the line $\mathcal{L}$, where it shows a discontinuity. Namely, it can be readily shown that (see Appendix B)
\begin{equation}
\Sigma(\omega+\epsilon)-\Sigma(\omega-\epsilon)=- 2 \pi ig_1(\omega)g_2(\omega) \rho(\omega)
\end{equation}
where $\omega=\omega(k)$ is a point of the contour $\mathcal{L}$, $\epsilon$ is a small complex number such that $\omega+\epsilon$ ($\omega-\epsilon$) falls outside (inside) the loop $\mathcal{L}$, and
\begin{equation}
\rho (\omega)=\frac{\partial k}{\partial \omega}
\end{equation}
is the density of states. Note that the density of states $ \rho(\omega)$, as a function of the complex variable $\omega$, is a continuous and single-valued function along the curve $\mathcal{L}$, however since $\oint_{\mathcal{L}} \rho ( \omega) d \omega=\oint_{\mathcal{L}}  ( d k / d \omega) d \omega=2 \pi$, $\rho(\omega)$ is not holomorphic inside the loop $\mathcal{L}$ and must show one (or more) poles or a set of branch points. Such singularities are the extension to complex crystals of van Hove singularities in ordinary (Hermitian) crystals, and correspond to the saddle points $\omega(k_s)$ defined by Eq.(7). 
Since the self-energy $\Sigma(\omega)$ is discontinuous as $\omega$ approaches the loop $\mathcal{L}$ from the inside or the outside, the curve $\mathcal{L}$ is a branch cut of $\hat{c}_a(\omega)$. 
Indicating by $\Sigma^{II}(\omega)$ and $\hat{c}_a^{II}(\omega)$ the analytic continuations of $\Sigma(\omega)$ and $\hat{c}_a(\omega)$ from the outside to the inside of the closed loop $\mathcal{L}$, the Bromwich path $\mathcal{B}$ can be deformed by crossing the loop $\mathcal{L}$. In doing so,  the Fourier-Laplace transform $\hat{c}_{a}(\omega)$ inside the loop $\mathcal{L}$ should be replaced by its analytic continuation $\hat{c}_a^{II}(\omega)$. Note that, according to Eq.(17), inside the loop $\mathcal{L}$ one has
\begin{eqnarray}
\Sigma^{II}(\omega) & = & \Sigma(\omega)-2 \pi i g_1(\omega) g_2(\omega) \rho (\omega) \\
\hat{c}_a^{II}(\omega) & = & \frac{1}{\omega-\omega_a-\Sigma(\omega)+2 \pi i g_1(\omega) g_2(\omega) \rho (\omega)}.
\end{eqnarray}
Therefore $\hat{c}_a^{II}(\omega)$ shows a set of branch points inside the loop $\mathcal{L}$ at the complex frequencies $\omega_s=\omega(k_s)$ defined by the saddle points (7) \cite{note}. To properly deform the Bromwich path $\mathcal{B}$ inside the loop $\mathcal{L}$, suitable branch cuts that connect the branch points should be therefore considered.\\ 
To further proceed in the analysis, let us assume the {\it weak coupling limit} $g_1g_2 \rightarrow 0$ and that $\omega=\omega_a$ is not a singularity (i.e. a pole or branch point) of $\rho(\omega)$, and let us distinguish two cases.\\
{\it First case: $\omega_a$ falls outside the loop $\mathcal{L}$.} In this case $\hat{c}_{a}(\omega)$ shows a pole at a frequency  $\omega=\omega_p$ close to $\omega_a$, i.e. outside the loop $\mathcal{L}$. The pole is found as a root of the equation
\begin{equation}
\omega_p-\omega_a-\Sigma(\omega_p)=0
\end{equation}
which in the weak coupling limit $g_1g_2 \rightarrow 0$ reads
\begin{equation}
\omega_p \simeq \omega_a +\Sigma(\omega_a).
\end{equation}
In this case the Bromwich path can be deformed as shown in Fig.2(b), where the dashed curves refer to the contour integrals of $\hat{c}_a^{II}(\omega)$ around the branch points $\omega_s$ internal to $\mathcal{L}$ whereas the solid circle is the contour integral of $\hat{c}_a(\omega)$ around the simple pole $\omega=\omega_p$. This yields
\begin{equation}
c_a(t)= \sqrt{\mathcal{Z}} \exp(-i \omega_p t) +c_a^{(cut)}(t)
\end{equation}
where $\sqrt{\mathcal{Z}} \simeq 1+ (d \Sigma / d \omega)_{\omega_a}$ is the residue of $\hat{c}_a(\omega)$ at the pole $\omega=\omega_p$, whereas $c_a^{(cut)}(t)$ is the contribution that arises from the branch cut integrals.\\
{\it Second case: $\omega_a$ falls inside the loop $\mathcal{L}$.} In this case $\hat{c}_{a}(\omega)$ shows a pole at a frequency  $\omega=\omega_p$ close to $\omega_a$, i.e. inside the loop $\mathcal{L}$. The Bromwich path can be deformed as in Fig.2(c), where the dashed curves refer to the contour integrals of $\hat{c}_a^{II}(\omega)$ around the branch points $\omega_s$ internal to $\mathcal{L}$ whereas the dashed circle is the contour integral of $\hat{c}_a^{II}(\omega)$ around the simple pole $\omega=\omega_p$. Therefore, the behavior of $c_a(t)$ is again given by Eq.(23), but with $\Sigma(\omega_a)$ replaced by $\Sigma^{II}(\omega_a)$. In particular, the pole $\omega_p$ is now given by [compare with Eq.(22)].
\begin{equation}
\omega_p \simeq \omega_a +\Sigma(\omega_a)-2 \pi i g_1(\omega_a) g_2 (\omega_a) \rho(\omega_a).
\end{equation}
Equation (23), together with Eqs.(22) and (24), are the main result of the present section and extend to the non-Hermitian continuum the general quantum mechanical decay law of the corresponding Hermitian problem \cite{r14}. A rich dynamical behavior can be envisaged depending on the values of the self-energy and density of states at $\omega=\omega_a$, and on the long-time behavior of the branch cut contribution.  As a general result (see Appendix C), it can be shown that if the unstable continuum is convectively (but not absolutely) unstable, the branch cut term $c_a^{(cut)}(t)$ is {\it decaying} at long times, whereas if the unstable continuum is absolutely unstable the branch cut term $c_a^{(cut)}(t)$ shows a secular growth, namely one has
\begin{equation}
 |c_a^{(cut)}(t)| \sim \frac{1}{ t^{1+\nu} } \exp [{\rm Im}(\omega_s)t]
 \end{equation}
as $t \rightarrow \infty$, where $\omega_s=\omega(k_s)$ is the complex energy of the most unstable saddle point [Eq.(7)] and $\nu>0$ is the power exponent of the branch point.\\ 
Some non-trivial dynamical behaviors can be predicted:\\
\\
(i) {\it Pseudo-Hermitian complete or fractional decay.} If the continuum is convectively (but not absolutely) unstable and $\Sigma(\omega_a)$ is real \cite{note2}, according to Eqs.(22) and (23) $c_a(t)$ shows a fractional decay when $\omega_a$ falls outside the energy loop $\mathcal{L}$ (the pole $\omega_p$ is real). On the other hand, if $\omega_a$ is embedded in the energy loop $\mathcal{L}$, from Eqs.(23) and (24) it follows that there is complete decay provided that the density of states $\rho(\omega_a)$ is real (the pole $\omega_p$ has a non vanishing negative imaginary part). Such results indicate that the decay dynamics in the convectively unstable continuum is analogous to the one in an Hermitian continuum. Such a regime will be thus referred to as  {\it pseudo-Hermitian quantum decay}.
However, since the continuum is not Hermitian, it might happen that $\rho(\omega_a)$ is imaginary. In this case, according to  Eqs.(23) and (24) it follows that, even though the energy $\omega_a$ of the discrete state is embedded inside the energy loop $\mathcal{L}$ of the continuum, the decay {\it is not} complete because the pole $\omega_p$ turns out to be real. Such a result does not have any counterpart in the corresponding Hermitian quantum decay problem, since the decay is always complete whenever the discrete-continuum coupling is weak and the energy of the discrete state is embedded in the continuous spectrum \cite{note3}. An example of fractional decay when $\omega_a$ is embedded into the continuum loop $\mathcal{L}$ will be presented in the next section.\\
\\
(ii) {\it Pseudo-exponential unstable growth}. If the continuum is absolutely unstable and the pole $\omega_p$ is real (this occurs, for example, whenever $\Sigma(\omega_a)$ is real \cite{note2} and $\omega_a$ is not embedded in the loop $\mathcal{L}$), $c_a(t)$ shows a secular growth which arises from the branch cut contribution solely [Eq.(23)]. According to Eq.(24), the unstable growth is not a pure exponential one, as one might expect at first sight, rather the exponential growing term is multiplied by an algebraic decaying term. \par
The above mentioned dynamical effects will be discussed in details in the next section by considering an exactly-solvable unstable tight-binding continuum.

\section{Quantum decay in a tight-binding continuum with asymmetric hopping rates}
Let us specialize the general results  obtained in the previous section considering in detail an exactly-solvable example. We consider a tight-binding lattice with asymmetric hopping rates $\kappa_1$ and $\kappa_2$ in the nearest-neighbor approximation \cite{uff1,uff2}, and assume that the discrete state $| a \rangle$ is coupled to the Wannier state $|0 \rangle$ of the lattice with an Hermitian hopping rate $\sigma$; see Fig.3(a).
The coupled equations for the site occupation amplitudes read
\begin{eqnarray}
i \frac{dc_n}{dt} & = &  \kappa_1 c_{n+1}+\kappa_2 c_{n-1}+ \sigma^* \delta_{n,0} c_a \\
i \frac{dc_a}{dt} & = &  \omega_a c_a + \sigma c_0. 
\end{eqnarray}
For the sake of definiteness, we will assume $\kappa_1>0$, whereas $\kappa_2$ can be either positive, negative or vanishing. The limiting case of an Hermitian continuum is obtained for $\kappa_2=\kappa_1$.
The dispersion relation $\omega(k)$ of the tight-binding lattice reads  [Fig.3(b)]
\begin{equation}
\omega(k)=\Delta_1 \cos k+ i \Delta_2 \sin k
\end{equation}
where we have set
\begin{equation}
\Delta_1= \kappa_1+ \kappa_2 \; \; , \;\;\; \Delta_2=\kappa_1-\kappa_2.
\end{equation}
Note that one has $\Delta_1>\Delta_2$ for $\kappa_2>0$, whereas $\Delta_2>\Delta_1$ for $\kappa_2<0$, with the limiting case $\Delta_1=\Delta_2$ for $\kappa_2=0$. Note also that,
as $k$ spans the Brillouin zone from $k= -\pi$ to $k=\pi$,  $\omega(k)$ describes an ellipse $\mathcal{L}$ in the complex energy plane, with the major axis oriented along the horizontal  (vertical) axis when $\Delta_1>\Delta_2$ ($\Delta_1<\Delta_2$); see Fig.3(c). In the limiting case $\Delta_2=\Delta_1$ the ellipse degenerates into a circle. The density of state $\rho(\omega)=(\partial k / \partial \omega)$ and the spectral coupling functions $g_{1,2}(\omega)$ can be readily calculated and read
\begin{equation}
g_1(k)=g_2^{*}(k)=\frac{\sigma}{\sqrt{2 \pi}}
\end{equation}
\begin{equation}
\rho(\omega)=\frac{1}{\sqrt{\Gamma^2-\omega^2}}
\end{equation}
where we have set
\begin{equation}
\Gamma^2 \equiv \Delta_1^2-\Delta_2^2.
\end{equation}
Note that there are two saddle points at frequencies $\omega_{s1}=-\Gamma$  and $\omega_{s2}=\Gamma$, internal to the ellipse $\mathcal{L}$, which correspond to the singularities of the density of states $\rho(\omega)$ of order $\nu= 1/2$. For $\Delta_1>\Delta_2$  (i.e. $\kappa_2>0$), the saddle points lie on the real energy axis and thus the continuum is convectively unstable, whereas for $\Delta_1<\Delta_2$  (i.e. $\kappa_2<0$) one of the two saddle points has a positive imaginary part, indicating that the continuum is absolutely unstable. At $\Delta_2=\Delta_1$ (i.e. $\kappa_2=0$) the two saddle points coalesce and $\rho(\omega)$ shows a simple pole (rather than two branch points) at $\omega=0$; see Fig.3(c).\\
The self-energy $\Sigma(\omega)$ can be computed in an exact form using Eqs.(16), (28) and (30), and reads
\begin{equation}
\Sigma(\omega)= \left\{
\begin{array}{cc}
0 & {\rm \;\;\;\; \omega \; inside \; the \; ellipse \; \mathcal{L}} \\
-\frac{i |\sigma|^2}{\sqrt{\Gamma^2-\omega^2}} &  {\rm \;\;\;\; \omega \; outside \; the \; ellipse \; \mathcal{L}}
\end{array}
\right.
\end{equation}
Note that $\Sigma(\omega)$ is discontinuous along the line $\mathcal{L}$, and the discontinuity is related to the spectral functions $g_{1,2}(\omega)$ and density of states $\rho(\omega)$ according to Eq.(17). The analytic continuation $\Sigma^{II}(\omega)$ of $\Sigma(\omega)$ inside the ellipse $\mathcal{L}$ shows two branch points at $\omega=\omega_{s1}=-\Gamma$ and $\omega=\omega_{s2}=\Gamma$, which should be connected by a branch cut; see Fig.3(c). The Fourier-Laplace transform of $\hat{c}_{a}(t)$, analytically continued inside the ellipse $\mathcal{L}$, is then given by [Eq.(15)]
\begin{equation}
\hat{c}_a^{II}(\omega)= \frac{ \sqrt{\Gamma^2-\omega^2} }{i |\sigma|^2+(\omega-\omega_a) \sqrt{\Gamma^2 - \omega^2}}
\end{equation}
The pole $\omega=\omega_p$ of $\hat{c}_a^{II}(\omega)$ is found as the root of the equation
\begin{equation}
\Gamma^2-\omega^2=-\frac{| \sigma |^4}{(\omega-\omega_a)^2}
\end{equation}
which for $\sigma \rightarrow 0$ and $\omega_a$ far from $\pm \Gamma$ is approximately given by Eq.(24).
To determine the temporal evolution of $c_a(t)$, we should distinguish the three cases shown in Fig.3(c).\par

 \begin{figure*}
\includegraphics[scale=0.3]{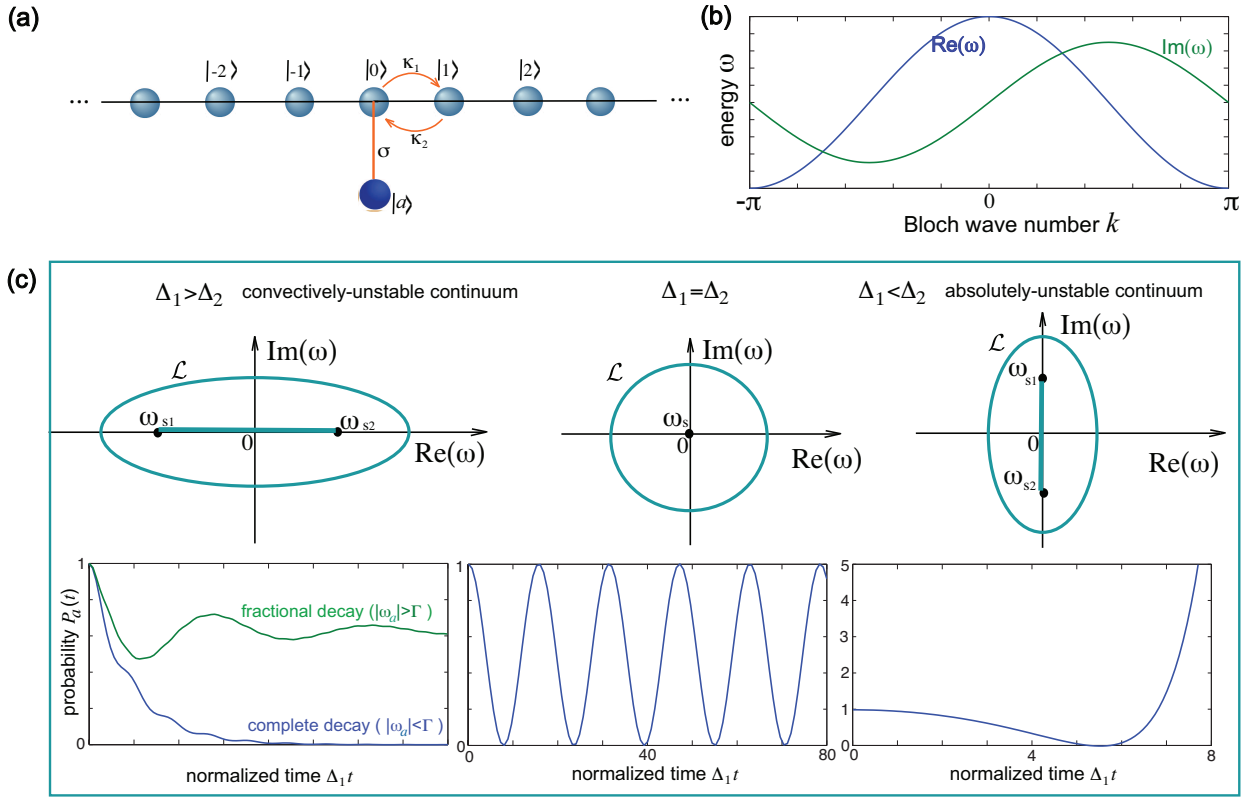}
\caption{(Color online) (a) Schematic of a discrete state $|a \rangle$ side coupled to the Wannier state $|0 \rangle$ of a tight-binding linear lattice with asymmetric hopping rates $\kappa_1$ and $\kappa_2$. (b) Behavior of the real and imaginary parts of the band dispersion relation $\omega(k)$ [Eq.(28)]. (c) Different dynamical behaviors corresponding to $\Delta_1 > \Delta_2$ (left panels), $\Delta_1=\Delta_2$ (central panels), and $\Delta_1<\Delta_2$ (right panels). For $\Delta_1> \Delta_2$, the continuum is convectively unstable, the density of states shows two branch points $\omega_{s1,2}$ on the real axis inside the ellipse $\mathcal{L}$, and $c_a(t)$ shows a complete or fractional decay, depending on whether $|\omega_a|$ is smaller or larger than $\Gamma= \sqrt{\Delta_1^2-\Delta_2^2}$.  For $\Delta_2> \Delta_1$, the continuum is absolutely unstable,  the density of states shows two branch points $\omega_{s1,2}$ on the imaginary axis inside the ellipse $\mathcal{L}$, and $c_a(t)$ shows a secular pseudo-exponential growth that arises from the brand cut contribution. For $\Delta_1=\Delta_2$ $c_a(t)$ shows an oscillatory (Rabi-like) behavior, regardless of the value of $\omega_a$ and coupling strength $|\sigma|$. The lower panels in (c) show the numerically-computed evolution of $P_a(t)$ in the three different regimes. Parameter values are as follows. Left panel:  $\Delta_2/ \Delta_1=0.7$, $\sigma / \Delta_1=0.2$, $\omega_a / \Delta_1=0$ (complete decay), $\omega_a / \Delta_1=0.8$ (fractional decay). Central panel: $\Delta_2/\Delta_1=1$, $\sigma / \Delta_1=0.2$, $\omega_a / \Delta_1=0$. Right panel: $\Delta_2/\Delta_1=1.2$, $\sigma / \Delta_1=0.2$, $\omega_a / \Delta_1=0$.}
\end{figure*}

{\it First case: $\Delta_1>\Delta_2$.} In this case the continuum is convectively unstable, the two branch points $\omega_{s1}=-\Gamma$ and $\omega_{s2}=\Gamma$ lie on the real axis and the cut contribution $c_a^{(cut)}(t)=(1/ 2 \pi) \int_{cut} d \omega \hat{c}^{II}_a(\omega) \exp(-i \omega t)$ decays algebraically as $t \rightarrow \infty$ according to Eq.(25) with $\nu= 1/2$. Therefore the non-decaying term of $c_a(t)$ comes from the pole contribution at $\omega=\omega_p$ [the first term on the right hand side of Eq.(23)]. Let us first consider the case where $\omega_a$ is inside the continuum loop $\mathcal{L}$. From Eqs.(24), (31) and (33) the pole $\omega_p$ is given by \cite{note4}
\begin{equation}
\omega_p \simeq \omega_a -i \frac{|\sigma|^2}{\sqrt{\Gamma^2-\omega_a^2}}
\end{equation}
Note that, for $|\omega_a| < \Gamma$ the pole has a non vanishing and negative imaginary part, i.e. there is a complete decay, $c_a(t) \rightarrow 0$ as $t \rightarrow \infty$; see left panel at the bottom in Fig.3(c). However, when $\omega_a$ is still inside the ellipse $\mathcal{L}$ but $\Gamma < |\omega_a| < \Delta_1$, the pole $\omega_p$ is real and thus the decay is fractional (limited), in spite the energy $\omega_a$ of the discrete state is embedded inside the continuum loop. Such a result does not have any counterpart in ordinary Hermitian quantum decay, where the decay is always complete when the energy of the discrete state is embedded within the continuum (except for special energy values corresponding to so-called bound states in the continuum or when $\omega_a$ is close to the edges of the continuum).  Let us now consider the case where $\omega_a$ is outside the continuum loop $\mathcal{L}$. From Eqs.(22) and (33) the pole $\omega_p$ is now given by 
\begin{equation}
\omega_p \simeq \omega_a + \frac{|\sigma|^2}{\sqrt{\omega_a^2-\Gamma^2}}
\end{equation}
which is real. Therefore the amplitude $c_a(t)$ does not vanish as $t \rightarrow \infty$ and the decay is fractional [see left panel in Fig.3(c)].\par
{\it Second case: $\Delta_1=\Delta_2$.} In this case the two branch points $\omega_{s1}$ and $\omega_{s2}$ coalesce, yielding a pole at $\omega=0$ for $\Sigma^{II}(\omega)$. Therefore there is not any branch cut contribution to $c_a(t)$, and the exact behavior of $c_a(t)$ can be readily obtained by the inverse Laplace-Fourier transform
\begin{equation}
c_a(t)=\frac{1}{2 \pi } \int_{\mathcal{B}} d \omega \;  \frac{\omega}{-|\sigma|^2+\omega(\omega-\omega_a) } \exp(-i \omega t).
\end{equation}
The integral on the right-hand side of Eq.(38) can be computed by closing the Bromwich path $\mathcal{B}$ into the $\rm{Im}(\omega)<0$ half complex plane and using the residue theorem. 
Since the function under the sign of integral in Eq.(38) has two poles at $\omega= \Omega_{\pm}$, with
\begin{equation}
\Omega_{\pm}=\frac{\omega_a}{2} \pm \sqrt{\left( \frac{\omega_a}{2}\right)^2+|\sigma|^2}
\end{equation}
 from the residue theorem one readily obtains
 \begin{equation}
 c_a(t)=\frac{\Omega_+}{\Omega_+-\Omega_-} \exp(-i \Omega_+ t)-\frac{\Omega_-}{\Omega_+-\Omega_-} \exp( -i \Omega_-t).
 \end{equation}
This means that $P_a(t)=|c_a(t)|^2$ shows an oscillatory behavior at the frequency $(\Omega_+ - \Omega_-)= \sqrt{\omega_a^2+4 |\sigma|^2}$; see the central panel in Fig.3(c).\\
\\
{\it Third case: $\Delta_1<\Delta_2$.} In this case the continuum is absolutely unstable, the two branch points $\omega_{s1}=-\Gamma$ and $\omega_{s2}=\Gamma$ lie on the imaginary axis and the pole $\omega_p$ is real. From Eq.(23) it follows that the pole contribution to $c_a(t)$ [first term on the right hand side of Eq.(23)] does not decay but it is bounded, whereas the cut contribution $c_a^{(cut)}(t)$ is unbounded and shows a pseudo-exponential growth as $t \rightarrow \infty$ according to Eq.(25) with $\nu=1/2$ and ${\rm Im} (\omega_s)=|\Gamma|=\sqrt{\Delta_2^2-\Delta_1^2}$.\\
\\
Typical examples of the temporal evolution of $P_a(t)$ in the three above-mentioned cases, obtained by numerical simulations of the coupled equations (26) and (27), are shown in the bottom row of Fig.3(c). The pseudo-exponential amplification in the $\Delta_2>\Delta_1$ case is clearly shown in Fig.4. The figure depicts the numerically-computed behavior of $(1/t) {\rm log} [P_a(t)t^ \alpha]$ for a few values of $\alpha$, and the asymptotic limit ${\rm Im}(\omega_s)=\sqrt{\Delta_2^2-\Delta_1^2}$. Note that the asymptotic limit is at best approached for $\alpha=3/2$ (and not for $\alpha=0$), indicating that the amplification is not a pure exponential term.
 \begin{figure}
\includegraphics[scale=0.35]{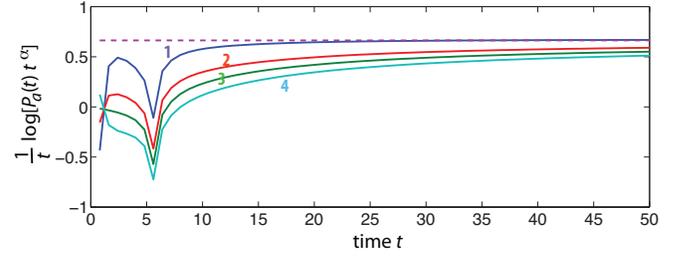}
\caption{(Color online) Numerically-computed behavior of $(1/t) {\rm log}[P_a(t) t^{\alpha}]$ for a few values of the power exponent $\alpha$ and for parameter values $\Delta_1=1$, $\Delta_2=1.2$, $\sigma=0.2$ and $\omega_a=0$. Curve 1: $\alpha=3/2$; curve 2: $\alpha=1/2$; curve 3: $\alpha=0$ (exponential amplification); curve 4: $\alpha=-1/2$. The dashed horizontal curve shows the asymptotic limit ${\rm Im}(\omega_s)=\sqrt{\Delta_2^2-\Delta_1^2} \simeq 0.66$.}
\end{figure}

The different behaviors of $P_a(t)$ found in the three above mentioned cases can be physically explained as follows. For $\Delta_1> \Delta_2$ the continuum is convective unstable, so that the initial excitation in the site $| a \rangle$ is continuously transferred to the Wannier site $|0 \rangle$ of the lattice and then amplified but convected away: this explains why full decay is still possible, despite  the excitation is getting amplified in the continuum. However, as compared to the Hermitian case the effective rate at which the excitation is convected away from the Wannier site $|0 \rangle$ is diminished because of the amplification that counteracts the convective motion. This is the reason why one can find a limited (fractional) decay when $|\omega_a|$ is smaller than $ \Delta_1$, i.e. even  when the discrete state is embedded into the energy loop $\mathcal{L}$ of the continuum. Such a behavior can be at best explained by observing that for $\Delta_1>\Delta_2$ the non-Hermitian decay problem given by Eqs.(26) and (27) can be reduced to an Hermitian problem by an 'imaginary' gauge transformation. In fact, after setting $\kappa_1=(\Gamma/2) \exp(h)$ and $\kappa_2=(\Gamma /2) \exp(-h)$, with $\Gamma^2=4 \kappa_1 \kappa_2= \Delta_1^2-\Delta_2^2$ and $h$ real, Eq.(26) takes the form
\begin{equation}
i \frac{dc_n}{dt}= \frac{\Gamma}{2} \left[ c_{n+1} \exp(h)+c_{n-1} \exp(-h) \right]+ \sigma \delta_{n,0}c_a.
\end{equation}
Note that in this form the tight-biniding continuum describes a lattice with (Hermitian) hopping rate $\Gamma/2$ between adjacent sites and with an 'imaginary' vector potential, which is accounted for by the complex Peierls' phase $ih$ \cite{uff1}. Such an imaginary vector potential can be eliminated by the 'imaginary' gauge transformation
\begin{equation}
c_n(t)=b_n(t) \exp(-hn).
\end{equation} 
Substitution of Eq.(42) into Eqs.(27) and (41) yields
\begin{eqnarray}
i \frac{db_n}{dt} & = &  \frac{\Gamma}{2} (b_{n+1}+b_{n-1}) + \sigma^* \delta_{n,0} c_a \\
i \frac{dc_a}{dt} & = &  \omega_a c_a + \sigma b_0. 
\end{eqnarray}
In their present form Eqs.(43) and (44) describe an {\it Hermitian} decay problem of the site $|a \rangle$ coupled to the Wannier site $|0 \rangle$ of an {\it Hermitian} tight-binding lattice (a quantum wire) with hopping rate $\Gamma /2$. Therefore, the decay dynamics in the original non-Hermitian problem when $\Delta_1 > \Delta_2$ is {\it pseudo-Hermitian}, i.e. it can be mapped into the quantum decay dynamics  of an effective Hermitian model. Note that the width $\Gamma$ of the effective tight-binding Hermitian lattice band is precisely that quantity that determines whether the quantum decay is complete ($|\omega_a|< \Gamma$, the discrete state $|a \rangle$ is embedded into the continuum) or fractional ($|\omega_a|> \Gamma$, the discrete state $|a \rangle$ is outside the continuum). Since $\Gamma < \Delta_1$, fractional decay can be observed when $\omega_a$ is embedded in the ellipse $\mathcal{L}$.\\ 
On the other hand, for $\Delta_2>\Delta_1$ the imaginary gauge transformation can not be applied and the decay problem is not pseudo-Hermitian. In this case basically the initial excitation of site $| a \rangle$ is transferred into the Wannier site $|0 \rangle$, however since in this case the continuum is absolutely unstable the excitation in $|0 \rangle$ is not advected away and undergoes a secular growth, which explains why $|a \rangle$ is secularly amplified.\\ The case $\Delta_1=\Delta_2$ is at the boundary between the convectively and absolutely unstable regimes, and can be analyzed directly in the Wannier basis representation using Eqs.(26) and (27). For $\Delta_1=\Delta_2$, one has $\kappa_2=0$ and thus Eqs.(26) and (27) read
\begin{eqnarray}
i \frac{dc_n}{dt} & = &  \kappa_1 c_{n+1}+ \sigma^* \delta_{n,0} c_a \\
i \frac{dc_a}{dt} & = &  \omega_a c_a + \sigma c_0. 
\end{eqnarray}
which should be integrated with the initial condition $c_a(0)=1$ and $c_n(0)=0$. It can be readily shown that the solution to the above equations with the given initial condition is given by
\begin{eqnarray}
c_n(t) & = & 0 \;\;\;\;\;\; n \leq -1 \\
c_0(t) & = & f_1(t) \\
c_n(t) & = & -i \kappa_1 \int_0^{t} dx c_{n-1}(x) \;\;\;\;  n \geq 1\\
c_a(t) & = & f_2(t)
\end{eqnarray}
where $f_{1,2}(t)$ satisfy the coupled equations
\begin{eqnarray}
i \frac{df_1}{dt} & = & \sigma^* f_2 \\
i \frac{df_2}{dt} & = & \sigma f_1 + \omega_a f_2
\end{eqnarray} 
with the initial conditions $f_1(0)=0$, $f_2(0)=1$. 
Such a solution shows that in the $\Delta_1=\Delta_2$ case the dynamics of the sites $|a \rangle$ and $|0 \rangle$ is decoupled from the other sites in the lattice, and thus Rabi-like oscillations are  observed, with the excitation being periodically transferred between sites $|a \rangle$ and $|0 \rangle$. The transfer in complete when $\omega_a=0$ (resonant Rabi oscillations), whereas it is incomplete for $\omega_a \neq 0$ (detuned Rabi oscillations). This result is in agreement with the exact solution given by Eqs.(39,40) and previously derived using the inverse Fourier-Laplace method.

\section{Conclusion}
The decay dynamics of a bound state coupled to a continuum is of major importance in different areas of physics. While in ordinary quantum mechanics the dynamics of the entire system must be unitary and the decay of the bound state is generally observed whenever its energy is embedded into the continuum, a richer dynamical behavior can be found when the Hamiltonian of the full system is allowed to be non-Hermitian and the dynamics described by a non-unitary operator.
In this work we have theoretically investigated the decay/amplification dynamics of a bound state weakly coupled to a {\it non-Hermitian unstable} continuum, i.e. a continuum containing non-normalizable (improper) eigenstates with positive imaginary part of the energy. A rich behavior has been disclosed, mainly depending on the absolute or convective nature of the instability of the continuum. In the former case the occupation amplitude of the discrete state grows secularly, however it may show a pseudo-exponential growth. In the latter case, full decay of the occupation amplitude of the discrete state can be observed, in spite the excitation transferred into the continuum is secularly amplified. Interestingly, limited (fractional) decay can be observed even when the energy of the discrete state is embedded within the energy loop of the continuum. The general analysis has been exemplified by considering in details the decay of a discrete state into a tight-binding lattice where instability arises from asymmetric hopping rates. Our results disclose novel dynamical features of discrete-continuum couplings in the non-Hermitian realm, and could stimulate further theoretical and experimental investigations. For example, it would be interesting to consider the impact of periodic "observations" of the system, i.e. Zeno and anti-Zeno dynamics, in the non-Hermitian realm.

\appendix

\section{Convectively and absolutely unstable tight-binding continua}
In this Appendix we briefly discuss the instability properties of the single-band tight-binding continuum, described by Eq.(3) with $\rho_n=0$. The most general solution to Eq.(3) is given by an arbitrary superposition of Bloch modes, namely
\begin{equation}
c_n(t)=\int_{-\pi}^{\pi}F(k) \exp[i k n-i \omega(k)t]
\end{equation}
where $\omega(k)$ is the dispersion relation, defined by Eq.(4) given in the text, and $F(k)$ is an arbitrary spectral function, that is determined by the initial condition $c_n(0)$ via the relation
\begin{equation}
F(k)=\frac{1}{2 \pi} \sum_{n=-\infty}^{\infty} c_n(0) \exp(-ikn).
\end{equation}
We wish to calculate the asymptotic behavior of $c_n(t)$ along the path $n=Vt$ as $t \rightarrow \infty$, i.e. of the function
\begin{equation}
c(t)=c_{n=Vt}(t)=\int_{-\pi}^{\pi} dk F(k) \exp[i k Vt -i \omega(k)t]
\end{equation}
where the index $n$ is here assumed to be a continuous variable.
Following the definition introduced in Ref.\cite{r30}, the tight-binding continuum is said to be {\it convectively unstable} if there exists a velocity $V \neq 0$ such that $c(t)$ is unbounded as $t \rightarrow \infty$, but $c_n(t) \rightarrow 0$ as $t \rightarrow \infty$ at any fixed lattice site $n$. The tight-binding continuum is said to be {\it absolutely unstable} if $c(t)$ is unbounded   as $t \rightarrow \infty$ for $V=0$. For the determination
of the asymptotic behavior of $c(t)$ we only need
to evaluate the integral on the right-hand side of Eq. (A3)
for those values of $k$ for which ${\rm Im}[\omega(k)]>0$, the other
modes giving no contribution (they are surely decaying). The
asymptotic behavior of $c(t)$ can be determined by the
saddle-point (or steepest descent) method \cite{r31,asym}.  A saddle point $k=k_s$ is found as the root of the equation
\begin{equation}
\left( \frac{d \omega}{dk} \right)_{k_s}=V.
\end{equation}
Indicating by $k=k_s$ the most unstable saddle point, i.e. with the highest imaginary part of the energy $\omega(k_s)$, and by  $n \geq 2$ its order, i.e. $\omega(k)=\omega(k_s)+V(k-k_s)+(d^n \omega /dk^n)_{k_s} (k-k_s )^n + o((k-k_s)^n)$, for $t  \rightarrow \infty$ one has 
\begin{eqnarray}
c(t) & \sim & 
\frac{F(k_s)}
{|t (d^n \omega/dk^n)_{k_s}|^{1/n}}
(n!)^{1/n} \Gamma \left( \frac{1}{n} \right) \nonumber \\
& \times &  \exp[itVk_s \pm i \pi /(2n)] \exp[-itE(k_s )].
\end{eqnarray}
From Eqs.(A4) and (A5) it readily follows that the continuum is absolutely unstable whenever the energy $\omega(k_s)$ at the saddle point $k=k_s$, satisfying Eq.(7) given in the text, has a positive imaginary part. It can be readily shown that, whenever the continuum is not absolutely unstable but $\omega(k)$ has an imaginary positive part for some wave numbers $k$, the continuum is convectively unstable. In fact, let us indicate by $k=k_0$ the maximum of the imaginary part of $\omega(k)$, and let us assume a drift velocity $V= {\rm Re}[\omega(k_0)]$. Then Eq.(A4) is satisfied for $k_s=k_0$. Since ${\rm Im}[\omega(k_0)]>0$, from Eq.(A5) it then follows that $c(t)$ is unbounded. 

\section{Discontinuity of the self-energy}
In this Appendix we prove Eq.(17) given in the text. To this aim, let us indicate by $\omega_0=\omega(k_0)$ a point on the contour $\mathcal{L}$ and by $\epsilon$ an infinitesimal  complex number such that the two points $\omega_0+\epsilon$ and   $\omega_0-\epsilon$ fall outside and inside the loop $\mathcal{L}$, respectively, and the segment connecting them is orthogonal to $\mathcal{L}$. Such a condition can be satisfied by assuming 
\begin{equation}
\epsilon=i \left( \frac{d \omega}{d k} \right)_{k_0} \delta
\end{equation}
with $\delta$ real and $\delta \rightarrow 0$.  From the definition of the self-energy [Eq.(16)] one readily obtains
\begin{eqnarray}
\Sigma(\omega_0+\epsilon)-\Sigma(\omega_0-\epsilon)=- 2 \epsilon  \\
\times  \int_{-\pi}^{\pi}dk \frac{g_1(k)g_2(k)}{[\omega_0+\epsilon-\omega(k)][\omega_0-\epsilon-\omega(k)]}. \nonumber
\end{eqnarray}
As $|\epsilon| \rightarrow 0$, the main contribution to the integral on the right hand side of Eq.(B2) comes from the wave numbers $k$ around $k=k_0$. This is because at $k=k_0$ the function under the sign of integral diverges as $\sim 1 / |\epsilon|^2$. Therefore we can set $\omega(k) \simeq \omega_0+( d \omega / dk)_{k_0} (k-k_0)$ in the denominator on the right hand side of Eq.(B2) and $g_1(k)g_2(k) \simeq g_1(k_0)g_2(k_0)$ in the numerator. After extending the integrals from $k=-\infty$ to $k=\infty$ one obtains
\begin{equation}
\Sigma(\omega_0+\epsilon)-\Sigma(\omega_0-\epsilon) \simeq -\frac{2 i \delta g_1(k_0) g_2(k_0) }{(d \omega / dk)_{k_0}} \int_{-\infty}^{\infty} \frac{dk}{\delta^2+(k-k_0)^2}
\end{equation}
which is exact in the $\delta \rightarrow 0$ limit. Taking into account that
\begin{equation}
\int_{-\infty}^{\infty} \frac{ dx}{\delta^2+x^2}=\frac{\pi}{\delta}
\end{equation}
one finally obtains
\begin{equation}
\Sigma(\omega_0+\epsilon)-\Sigma(\omega_0-\epsilon)= -\frac{2 \pi i  g_1(k_0) g_2(k_0) }{(d \omega / dk)_{k_0}}
\end{equation}
which is Eq.(17) given in the text once the density of states $\rho(\omega)=(dk/ d \omega)$ is introduced and the product $g_1g_2$ is written as a function of frequency (rather than wave number). 
 \begin{figure}
\includegraphics[scale=0.5]{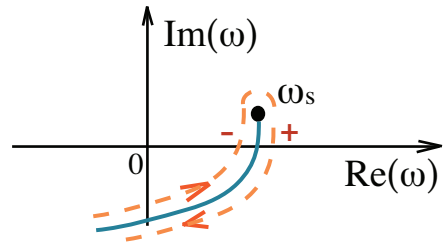}
\caption{(Color online) Schematic of the branch cut departing from a branch point $\omega=\omega_s$ inside the loop $\mathcal{L}$. The branch cut is depicted by the solid curve and parametrized by the equation $\omega=\omega_b(s)$, where $s$ is the arc length of the curve starting from the branch point. The dashed line shows the contour integral around the branch cut.}
\end{figure}
\section{Branch cut contribution}
In this Appendix we prove Eq.(25) given in the text. To this aim, let us consider a branch point at frequency $\omega=\omega_s$ inside the loop $\mathcal{L}$, and let us compute the contribution to the branch cut contour in the neighborhood of the branch point. The branch cut departing from $\omega=\omega_s$ is described by the curve $\omega_b=\omega_b(s)$, where $s$ is the curvilinear abscissa with $s=0$ at the branch point, i.e. $\omega_b(s=0)=\omega_s$. Without loss of generality, in the neighborhood of $\omega_s$ the branch cut may be chosen so as the imaginary part of $\omega_b(s)$ decreases as $s$ increases from zero, i.e. ${\rm Im}[\omega_b(s_2)]>{\rm Im}[\omega_b(s_1)]$ for $s_2>s_1$, with $(d \omega_b / ds )_{s=0}=-iR$ imaginary and $R>0$. As shown in Fig.5, this means that the branch point $\omega_s$ is at the top of the branch cut in the complex energy plane. 
Let us now calculate the contribution of the branch cut contour to $c_a(t)$, i.e. the integral
\begin{eqnarray}
c_a^{(cut)}(t) & = & \frac{1}{2 \pi} \int_{cut} d \omega \hat{c}_a^{II}(\omega) \exp(-i \omega t)=  \frac{1}{2 \pi} \exp(-i \omega_s t) \nonumber \\
&   \times & \int_{0}^{s_f} ds \frac{d \omega_b}{ds}  F(s) \exp\{-i [\omega_b(s)-\omega_s] t\} 
\end{eqnarray}
where $F(s)$ is the difference between $c_a^{II}(\omega)$ evaluated form the right(+) and from the left(-) sides of the branch cut (see Fig.5), and $s_f$ is the arc length of the branch cut. 
To evaluate the integral on the right hand side of Eq.(C1) in the $t \rightarrow \infty$ limit, let us note that, since ${\rm Im}[\omega_b(s)-\omega_s]$ is a negative decreasing function of $s$, vanishing at $s=0$, the main contribution to the integral arises for $s$ close to zero. Moreover, since $\omega=\omega_s$ is a branch point of $c_a^{II}(\omega)$ arising from a singularity $\rho(\omega) \sim 1/ (\omega-\omega_s)^ \nu$ in the density of states, from Eq.(20) it follows that in the neighborhood of $\omega=\omega_s$ one has 
\begin{equation}
\hat{c}_{a}^{II}(\omega) \sim A (\omega-\omega_s)^{\nu}
\end{equation}
where $\nu>0$ is the of order of the branch point and $A$ a constant. The difference $F(s)$ between $c_a^{II}(\omega)$ evaluated form the right (+) and from the left (-) sides of the branch cut 
is dictated by the change of $ (\omega-\omega_s)^{\nu}$ as $ \omega-\omega_s=\epsilon \exp(i \varphi)$ describes a circle around the branch cut, from $\varphi=0$ to $\varphi=2 \pi$. Hence one obtains
\begin{equation}
F(s) \sim F_0  [\omega_b(s)-\omega_s]^{\nu}
\end{equation}
where we have set $F_0=A [1-\exp( 2 \pi i \nu)]$.
In the asymptotic limit $t \rightarrow \infty$, we can assume in Eq.(C1) $\omega_b(s)-\omega_s \simeq d(\omega_b/ds)_{s=0} s=-iRs$, $F(s) \sim F_0  [\omega_b(s)-\omega_s]^{\nu} \sim F_0 (-iRs)^{\nu}$ and we may extend the integral to $\infty$. One obtains 
\begin{equation}
c_a^{(cut)}(t)  \simeq   \frac{F_0 (-i R)^{\nu+1}}{2 \pi} \exp(-i \omega_s t)
 \int_{0}^{\infty} ds s^{\nu} \exp(-Rst)
\end{equation}
as $t \rightarrow \infty$. Taking into account the definition of the Gamma function $\Gamma(z)=\int_0^{\infty}dx \;  x^{z-1} \exp(-x)$, after the change of variable $x=Rst$ in the integral on the right hand side of Eq.(C4) one finally obtains
\begin{equation}
c_a^{(cut)}(t)  \simeq   \frac{F_0  \Gamma( \nu+1)  \exp[-i \pi (\nu+1)/2]}{2 \pi} \frac{1}{t^{\nu+1}}\exp(-i \omega_s t).
\end{equation}
From Eq.(C5) is follows that $|c_a(t)| \sim t^{-(\nu+1)} \exp[{\rm Im}(\omega_s) t]$ as $t \rightarrow \infty$, which is Eq.(25) given in the text.


\begin{thebibliography}{99}

\bibitem{r1}
L. Fonda, G. C. Ghirardi, and A. Rimini, Rep. Prog.
Phys. {\bf 41}, 587 (1978).

\bibitem{r2}
H. Nakazato, M. Namiki, and S. Pascazio,  Int. J. Mod. Phys. B {\bf 10}, 247 (1996).

\bibitem{r3}
L. S. Schulman, Lecture Notes Phys. {\bf 734}, 107 (2008).

\bibitem{r4}
T. Gorin, T. Prosen, T.H. Seligman, and M. Znidaric, Phys. Rep. {\bf 425}, 33 (2006).

\bibitem{r5}
C. Gogolin and J. Eisert, Rep. Prog. Phys. {\bf 79}, 056001 (2016).

\bibitem{r6}
P.L. Knight, M.A. Lauder, and B.J. Dalton, Phys. Rep. {\bf 190}, 1 (1990).

\bibitem{r6bis}
N. Moiseyev, Phys. Rep, {\bf 302}, 211 (1998).

\bibitem{r6tris} 
T Petrosky, C.O. Ting, and S. Garmon,
Phys. Rev. Lett. {\bf 94}, 043601 (2005);  H. Nakamura, N. Hatano, S. Garmon, and T. Petrosky,
Phys. Rev. Lett. {\bf 99}, 210404 (2007).

\bibitem{r7}
A.G. Kofman, G. Kurizki, and B. Sherman, J.  Mod. Opt. {\bf 41},  353 (1994); 
P. Lambropoulos, G. M. Nikolopoulos, T. R. Nielsen, and S. Bay, Rep. Prog. Phys. {\bf 63}, 455 (2000).

\bibitem{r8}
S. Longhi, Phys. Rev. Lett. {\bf 97}, 110402 (2006); S. Longhi, Laser \& Photon. Rev. {\bf 3}, 243 (2009).

\bibitem{r9}
L.M. Krauss and J. Dent, Phys. Rev. Lett. {\bf 100}, 171301 (2008).

\bibitem{r10}
S. R. Wilkinson, C. F. Bharucha, M. C. Fischer, K.W. Madison, P. R. Morrow, Q. Niu, B. Sundaram, and M. G.
Raizen, Nature {\bf 387}, 575 (1997).

\bibitem{r11}
C. B. Chiu, B. Misra, and E. C. G. Sudarshan, Phys. Rev. D {\bf 16}, 520 (1977).

\bibitem{r12}
A. G. Kofman and G. Kurizki, Nature {\bf 405}, 546 (2000); A.G. Kofman and G. Kurizki,
Phys. Rev. Lett. {\bf 87},  270405 (2001).

\bibitem{r13}
M. C. Fischer, B. Gutierrez-Medina, and M. G. Raizen, Phys. Rev. Lett. {\bf 87}, 040402 (2001).

\bibitem{r14}
P. Facchi, H. Nakazato, and S. Pascazio, Phys. Rev. Lett. {\bf 86}, 2699 (2001).

\bibitem{r15}
J. M. Raimond, P. Facchi, B. Peaudecerf, S. Pascazio, C. Sayrin, I. Dotsenko, S. Gleyzes, M. Brune, and S. Haroche
Phys. Rev. A {\bf 86}, 032120 (2012).

\bibitem{r16}
Y.S. Patil, S. Chakram, and M. Vengalattore, Phys. Rev. Lett. {\bf 115}, 140402  (2015).

\bibitem{r17}
F. Dreisow, A. Szameit, M. Heinrich, T. Pertsch, S. Nolte, A. T\"{u}nnermann, and S Longhi, Phys. Rev. Lett. {\bf 101}, 143602 (2008);
P. Biagioni, G. Della Valle, M. Ornigotti, M. Finazzi, L. Duo, P. Laporta, and S. Longhi, Opt. Express {\bf 16}, 3762 (2008);
M.A. Porras, A. Luis, I. Gonzalo, and A.S. Sanz, Phys. Rev. A {\bf 84}, 052109 (2011).

\bibitem{r17bis}
J. Martorell, J. G. Muga, and D. W. L. Sprung, Phys. Rev. A {\bf 77}, 042719 (2008).

\bibitem{r18}
A. del Campo, F. Delgado, G. Garcia-Calderon, J. G. Muga, and M.G. Raizen, Phys. Rev. A {\bf 74}, 013605 (2006); 
A.U.J. Lode, A.I. Streltsov, O.E. Alon, H.-D. Meyer, and L.S. Cederbaum,
J. Phys. B {\bf 42}, 044018 (2009); T. Taniguchi and S.I. Sawada, Phys. Rev. E {\bf 83}, 026208 (2011); 
A. del Campo, Phys. Rev. A {\bf 84}, 012113 (2011);  G. Garcia-Calderon and L. G. Mendoza-Luna, Phys. Rev. A {\bf 84}, 032106 (2011); M. Rontani, Phys. Rev. Lett. {\bf 108}, 115302 (2012);
M. Pons, D. Sokolovski, and A. del Campo, Phys. Rev. A {\bf 85}, 022107 (2012); S. Longhi and G. Della Valle, Phys. Rev. A {\bf 86}, 012112 (2012); 
A. Del Campo, New J. Phys. {\bf 16}, 015014 (2016).

\bibitem{r19}
A. Crespi, L. Sansoni, G. Della Valle, A. Ciamei, R. Ramponi, F. Sciarrino, P. Mataloni, S. Longhi, and R. Osellame, Phys. Rev. Lett. {\bf 114}, 090201 (2015).

\bibitem{r20}
P.M. Preiss, R. Ma, M.E. Tai, A. Lukin, M. Rispoli, P. Zupancic, Y. Lahini, R. Islam, and M. Greiner,  Science {\bf 347}, 1229 (2015).

\bibitem{r21}
N. Moiseyev, {\it Non-Hermitian Quantum Mechanics} (Cambridge University Press, Cambridge, 2011)

\bibitem{r22}
C. Bender, Rep. Prog. Phys. {\bf 70}, 947 (2007). 

\bibitem{r23}
S. Longhi, Phys. Rev. A {\bf 82}, 031801 (2010);
S. Zhang, Z. Ye, Y.Wang, Y. Park, G. Bartal, M. Mrejen, X. Yin,
and X. Zhang, Phys. Rev. Lett. {\bf 109}, 193902 (2012); P. Ginzburg
F. J. Rodriguez-Fortuno, A. Martinez, and A. V. Zayats, Nano
Lett. {\bf 12}, 6309 (2012); G. Castaldi, S. Savoia, V. Galdi, A. Alu,
and N. Engheta, Phys. Rev. Lett. {\bf 110}, 173901 (2013);
 L. Feng, Y.-L. Xu, W. S. Fegadolli, M.-H. Lu, J. E. B. Oliveira, V. R.
Almeida, Y.-F. Chen, and A. Scherer, Nature Mat. {\bf 12}, 108 (2013);
F. Nazari, N. Bender, H. Ramezani, M. K. Moravvej-Farshi, D. N.
Christodoulides, and T. Kottos, Opt. Express {\bf 22}, 9575 (2014);
 B. Peng, S. K. Ozdemir, F. Lei, F. Monifi, M. Gianfreda, G. L. Long, S.
Fan, F. Nori, C. M. Bender, and L. Yang, Nature Phys. {\bf 10}, 394 (2014);
 H. Hodaei, M. A. Miri, M. Heinrich, D. N. Christodoulides, and M.
Khajavikhan, Science {\bf 346}, 975 (2014); L. Feng, Z. J. Wong, R. M. Ma, Y. Wang, and X. Zhang, Science {\bf 346},
972 (2014); S. Longhi, Opt. Lett. {\bf 41}, 1897 (2016).

\bibitem{r24}
S. Longhi, Phys. Rev. A {\bf 74}, 063826 (2006); S. Longhi, Phys. Rev. B {\bf 80}, 165125  (2009);
A. Regensburger, M.-A. Miri, C. Bersch, J. N\"{a}ger, G.
Onishchukov, D. N. Christodoulides, and U. Peschel, Phys. Rev. Lett. {\bf 110}, 223902 (2013);
S. Longhi, Opt. Lett. {\bf 39}, 1697 (2014); S. Longhi and G. Della Valle, Phys. Rev. A {\bf 89}, 052132 (2014).

\bibitem{r25}
M. V. Berry and D. H. J. ODell, J. Phys. A {\bf 31}, 2093 (1998);
M. K. Oberthaler, R. Abfalterer, S. Bernet, J. Schmiedmayer,
and A. Zeilinger, Phys. Rev. Lett. {\bf 77}, 4980 (1996); C. Keller,
M. K. Oberthaler, R. Abfalterer, S. Bernet, J. Schmiedmayer,
and A. Zeilinger, Phys. Rev. lett. {\bf 79}, 3327 (1997); 
R. St\"{u}tzle, M. C. G\"{o}bel, T. H\"{o}rner, E. Kierig, I. Mourachko, M.
K. Oberthaler, M. A. Efremov, M. V. Fedorov, V. P. Yakovlev,
K. A. H. van Leeuwen, and W. P. Schleich, Phys. Rev. Lett. {\bf 95},
110405 (2005).

\bibitem{r26}
K. G. Makris, R. El-Ganainy, D. N. Christodoulides,
and Z. H. Musslimani, Phys. Rev. Lett. {\bf 100}, 103904
(2008); S. Longhi, Phys. Rev. Lett. {\bf 103}, 123601 (2009); S. Longhi, Phys. Rev. B
{\bf 80}, 235102 (2009); L. Feng, M. Ayache, J. Huang, Y.-L. Xu, M.-H. Lu, Y.-F. Chen,
Y. Fainman, and A. Scherer, Science {\bf 333}, 729 (2011); A. Regensburger, C. Bersch, M.-A. Miri, G. Onishchukov, D. N.
Christodoulides, and U. Peschel, Nature {\bf 488}, 167 (2012).

\bibitem{r27}
Z. Lin, H. Ramezani, T. Eichelkraut, T. Kottos, H. Cao, and
D. N. Christodoulides, Phys. Rev. Lett. {\bf 106}, 213901 (2011); S.
Longhi, J. Phys. A {\bf 44}, 485302 (2011); E.-M. Graefe and H. F. Jones,
Phys. Rev. A {\bf 84}, 013818 (2011); B. Midya, Phys. Rev. A {\bf 89}, 032116 (2014).

\bibitem{r28}
S. Longhi, Phys. Rev. A {\bf 90}, 043827 (2014).

\bibitem{r29}
S. Longhi, G. Della Valle, and K. Staliunas, Phys. Rev. A {\bf 84}, 042119 (2011). 

\bibitem{r30}
S. Longhi, Phys. Rev. A {\bf 88}, 052102 (2013).

\bibitem{r31}
P. Huerre and P. A.Monkewitz, Annu. Rev. Fluid Mech. {\bf 22}, 473
(1990); L. S. Hall and W. Heckrotte, Phys. Rev. {\bf 166}, 120 (1968); R. J.
Deissler, J. Stat. Phys. {\bf 40}, 371 (1985); Physica D (Amsterdam)
{\bf 56}, 303 (1992).

\bibitem{note}
Note that the branch points of $\hat{c}_a^{II}(\omega)$ inside the loop $\mathcal{L}$ are only the singularities $\omega_s=\omega(k_s)$ of the density of states $\rho(\omega)$ with the 
asymptotic behavior $\rho(\omega) \sim (\omega-\omega_s)^ {-\nu}$ near $\omega=\omega_s$, with $\nu>0$ a non-integer number. In fact, a pole of $\rho(\omega)$ ($ \nu$ integer) just corresponds to a zero (rather than a branch point) of $\hat{c}_a^{II}(\omega)$. 

\bibitem{note2}
$\Sigma(\omega_a)$ turns out to be real, for example, whenever the dispersion relation $\omega(k)$ satisfies the condition $\omega(-k)=\omega^*(k)$ and $g_1(k)=g_2^*(k)$ has a definite parity (either odd or even) for the inversion $k \rightarrow -k$.

\bibitem{note3}
We do not consider here the rather special case $g_1(\omega_a)g_2(\omega_a)=0$, which does not lead to complete decay because of the existence of a bound state in the continuum [see, for instance: M. Miyamoto, Phys. Rev. A {\bf 72}, 063405 (2005); S. Longhi. Eur. J. Phys. B {\bf 57}, 45 (2007)]. We also do not consider the very special case where the energy $\omega_a$ is at the edge of the continuum.

\bibitem{uff1}
N. Hatano and D. R. Nelson, Phys. Rev. Lett. {\bf 77}, 570 (1996).

\bibitem{uff2}
S. Longhi, Phys. Rev. A {\bf 82}, 032111 (2010); X. Z. Zhang and  Z. Song, Ann.  Phys. {\bf 339}, 109 (2013);  
S. Longhi, EPL {\bf 106}, 34001 (2014);
S. Longhi, Phys. Rev. A {\bf 92}, 042116 (2015); S. Longhi, D. Gatti, and G. Della Valle, Sci. Rep. {\bf 5}, 13376
(2015).

\bibitem{note4}
Such an expression holds when $\omega_a$ is sufficiently far from a singularity of the density of states $\rho(\omega)$, i.e. for $\omega_a$ far from $ \pm \Gamma$. As $\omega_a$ approaches $\pm \Gamma$ from either above or below, the exact algebraic equation (35) should be used to compute the pole $\omega_p$. 

\bibitem{asym}
F. W. J. Olver, {\it Asymptotics and Special Functions} (Academic, New York, 1974).

\bibitem{note5}
If there are mode saddle points, one has to consider the most unstable one, corresponding to the largest part of ${\rm Im} [\omega(k_s)]$.












\end{thebibliography}
\end{document}